\documentclass[final,1p,times,]{elsarticle}
\usepackage{ifpdf}
\usepackage[usenames]{color}
\usepackage{graphicx}
\usepackage{amssymb}
\usepackage{amsmath}
\usepackage{ulem}
\journal{Applied Sciences}
\begin{document}
\begin{frontmatter}
\title{Interplay between binary and three-body interactions and
enhancement of stability in trapless dipolar Bose-Einstein condensates}
\author{ Sabari Subramaniyan$^{1}$, Kishor Kumar Ramavarmaraja $^{2}$, Radha Ramaswamy $^{1}$* and Boris A Malomed$^{3,4}\dagger$}
\address{$^{1}$Centre for Nonlinear Science, Department of Physics, Government College for Women(A), Kumbakonam 612001, India\\
$^{2}$Department of Physics, Centre for Quantum Science, and Dodd-Walls Centre for Photonic and Quantum Technologies, University of Otago, Dunedin 9054, New Zealand.\\ 
$^{3}$Department of Physical Electronics, School of Electrical Engineering, Faculty of Engineering, and Center for Light-Matter Interaction, Tel Aviv University, P.O.B. 39040, Ramat Aviv, Tel Aviv, Israel \\
$^{4}$Instituto de Alta Investigaci\'{o}n, Universidad de Tarapac\'{a}, Casilla 7D, Arica, Chile.
\\$^*$vittal.cnls@gmail.com (R.R);\,\,$^\dagger$malomed@tauex.tau.ac.il (B.A.M)}
\date{\today}
\begin{abstract}
We investigate the nonlocal Gross-Pitaevskii (GP) equation with long-range
dipole-dipole and contact interactions (including binary and three-body
collisions). We address the impact of the three-body interaction on
stabilizing trapless dipolar Bose-Einstein condensates (BECs). It is found
that the dipolar BECs exhibit stability not only for the usual combination
of attractive binary and repulsive three-body interactions, but also for the
case when these terms have opposite signs. The trapless stability of the
dipolar BECs may be further enhanced by time-periodic modulation of the
three-body interaction imposed by means of Feshbach resonance. The results are produced analytically using the variational approach and confirmed by numerical simulations.
\end{abstract}


\end{frontmatter}

\date{\today}
\section{Introduction}
\label{sec1} 
The advent of Bose-Einstein condensates (BECs) in $^{52}$Cr~\cite{Koch:2008,Lahaye:2009}, $^{164}$Dy~\cite{Lu:2011,Youn:2010} and $^{168}$Er~\cite{Aikawa:2012} accompanied by long-range dipole-dipole (DD) interactions superimposed on the contact inter-atomic collisions has impacted the investigation of ultracold quantum gases~\cite{Baranov:2008}. The anisotropic character and long-range nature of DD interactions endows the dipolar BECs (DBECs) with several distinct features such as the subordination of stability on the trap geometry~\cite{Koch:2008,Lahaye:2009}, roton-maxon character of the excitation spectrum~\cite{Santos:2003,Goral:2002}, new dispersion relations for elementary excitations~\cite{Wilson:2010,Ticknor:2011}, novel quantum phases~\cite{Tieleman:2011,Zhou:2010}, explicit \cite{vor1,vor2} and hidden vortices~\cite{Sabari2017hv}, specific vortex-antivortex pairs~\cite{Sabari2018a}, anisotropic multidimensional solitons \cite{Tikhonenkov:2008,Koberle}, quantum droplets stabilized by beyond-mean-field effects \cite{Pfau1,Pfau3,Santos1}, etc. The above phenomena arise due to the interplay between the contact \textit{s}-wave interactions and the dipolar attraction or repulsion \cite{competing}. Tuning contact interactions by means of the Feshbach resonance is an important tool in analyzing the properties of DBECs \cite{Santos:2003,Goral:2002}.

The theoretical description of a dilute weakly interacting DBEC is based on
the Gross-Pitaevskii (GP) equation with the nonlocal DD-interaction term
\cite{Koch:2008,Lahaye:2009,Muruganandam:2012,Lahaye:2008,RKK:2015}. In particular, the combination of local and nonlocal terms in the GP
equation can support various species of bright and dark matter-wave
solitons. In the alkali BECs, bright solitons exist when the negative
(attractive) contact interaction exactly balances the dispersion
(kinetic-energy) term~\cite{Strecker:2002,Khaykovich:1995}. In
DBEC, the nonlocal DD interaction term may reinforce local ones originating
from the \textit{s}-wave contact interaction. DD interactions have caught the limit with the atomic condensates of lanthanide series
elements such as $^{168}$Er and $^{164}$Dy. Long-range interactions also
play a crucially important role in optical of nonlocal media, where they
induce modulational instability, solitons, and vortices \cite%
{Krolikowski:2001,Bang:2002}. The DD interactions in BECs can support bright
solitons even for the positive (repulsive) contact interaction~\cite{Muruganandam:2012}.

Following the scheme of the stabilization of the inverted (Kapitza) pendulum~%
\cite{Landau:1960}, scenarios for stabilization of two-dimensional (2D)
optical \cite{Towers2002} and matter-waves \cite%
{Abdullaev2003,Saito2003,Adhikari2004} by means of the \textquotedblleft
nonlinearity management" \cite{book}, i.e., the cubic nonlinearity
periodically switching between self-attraction and repulsion, have been
elaborated. This concept has been subsequently applied to 3D vortex solitons~\cite{Abdullaev2003,Adhikari2004} and extended to the model containing the three-body interaction~\cite{Sabari2010} and quantum fluctuations~\cite{Sabari2017}. The stabilization of self-repulsive BECs with periodically varying higher-order interactions has been
explored recently~\cite{Tamil}. The investigation of spatial and temporally varying nonlinearities has also drawn considerable interest in optics~\cite{Kivshar,Towers2002,Kartashov,Zeng}
and other BEC settings \cite{Sakaguchi,Abdullaev,Abdullaev2003a,Garcia,Pacciani,Belmonte1,Deng,Sabari2015,sabari2019}

The above investigations addressed systems with local interactions. Recently, the
possibility of the stabilization of DBECs by temporal modulation of the
contact interaction~was demonstrated too \cite{Sabari2018b}. In the latter
case, a potential minima necessary for self-trapping is found to be  absent without the
time-periodic modulation of the two- and three-body interaction. The current work
primarily addresses this situation, attempting to stabilize DBECs by adding
three-body terms to the binary interaction, cf. Refs.~\cite%
{Keltoum:2019,Boudjemaa:2018,Blakie:2016,Xi:2016,Bisset:2015,Lu:2015}. As a
result, the present work offers a simple protocol to stabilize
multidimensional trapless DBECs.

The organization of the paper is as follows. In Section~\ref{sec2}, we set
the mean-field model based on the nonlocal GP equation. Then, we introduce
the variational method in Section~\ref{sec3} and demonstrate the
stabilization of a trapless DBEC without time-periodic nodulation of the
contact interaction. Numerical results produced with the help of 
Runge-Kutta (RK4) method are also presented in Section~\ref{sec3}. In
Section~\ref{sec4}, we report numerical results for the 3D time-dependent GP
equation simulated by means of the Crank-Nicholson method. Concluding
remarks are presented in Section~\ref{sec5}.\newline

\section{The model}

\label{sec2}
At ultra-low temperatures, a DBEC with two-body, three-body and nonlocal DD interactions can be described by the following time-dependent GP equation~\cite{Koch:2008,Lahaye:2009,Muruganandam:2012,Lahaye:2008,RKK:2015,3body1}:

\begin{equation}
\begin{split}
\mathrm{i}\hbar \frac{\partial \phi ({\mathbf{r}},t)}{\partial t}=&\Big[-\frac{\hbar ^{2}}{2m}\nabla ^{2}+V({\mathbf{r}})+\frac{4\pi \hbar ^{2}a(t)N}{m}\left\vert \phi ({\mathbf{r}},t)\right\vert ^{2} +N\int U_{\mathrm{dd}}({\mathbf{r}}-{\mathbf{r}}^{\prime })\left\vert \phi ({\mathbf{r}}^{\prime },t)\right\vert ^{2}d{\mathbf{r}}^{\prime }\\ & +\eta(t) N^{2}\left\vert \phi ({\mathbf{r}},t)\right\vert ^{4}\Big]\phi ({\mathbf{r}},t),
\end{split}\label{dgpe1}
\end{equation}%
where $\phi ({\mathbf{r}},t)$ refers to the condensate wave function, $N$ the number of particles with magnetic dipole moment $\mu $, $m$ the mass, $\nabla^2$ the Laplacian operator, $\hbar$ the reduced Planck constant, $V({\mathbf{r}})=\frac{1}{2}m\left( \omega _{x}^{2}x^{2}+\omega_{y}^{2}y^{2}+\omega _{z}^{2}z^{2}\right) $ is the trapping potential with frequencies $\omega _{x},\omega _{y}$ and $\omega _{z}$, and $a(t)$ is the time-modulated atomic scattering length. The potential of the dipolar interaction for magnetic dipoles is $U_{\mathrm{dd}}(\mathbf{R})=\frac{\mu _0 \mu^2}{4\pi }\frac{1-3\cos^{2}\theta}{|\mathbf{R}|^3}$, where $\mathbf{R=r-r^{\prime }}$ determines the relative position of dipoles and $%
\theta $ is the angle between $\mathbf{R}$ and the vertical ($z$)
polarization direction, $\mu _{0}$ is the free-space permeability~\cite%
{Muruganandam:2012}. 
The parameter $\eta(t)$ is the strength of the three-body interaction
which depends on both the s-wave scattering length $a(t)$ and the
universal constants $d_1,d_2,a_0$ and $s_0$~\cite{3body1}. This parameter reads $\eta(t)=\frac{12\pi\hbar^2a(t)^2}{m}\left(d_1+d_2\,tan\left[s_0\,ln\frac{|a(t)|}{|a_0|}+\frac{\pi}{2}\right]\right)$ or $\eta(t)=\frac{12\pi\hbar^2a(t)^2}{m}\left(d_1-d_2\,cot\left[s_0\,ln\frac{|a(t)|}{|a_0|}\right]\right)$, where the numerical values of the universal constants are given in Refs.~\cite{3body1,3body2}. 
However, the strength of the three-body interaction is very small when compared with strength of the two-body interaction~\cite{Gammal2000}.  In some other regimes, the three-body interaction may  become a  dominant factor under special conditions as in  the cases of higher densities and Effimov resonance~\cite{3body1}. 
The normalization is
\begin{equation}
\int d\mathbf{r}|\phi ({\mathbf{r}},t)|^{2}=1.  
\label{N}
\end{equation}%
To compare the dipolar and contact interactions, it is often useful to
introduce the length scale $a_{\mathrm{dd}}\equiv \mu _{0}\mu%
^{2}m/(12\pi \hbar ^{2})$, whose experimental values for $^{52}$Cr, $^{164}$%
Er and $^{168}$Dy condensates are $16a_{0}$, $66a_{0}$ and $131a_{0}$,
respectively, $a_{0}$ being the Bohr radius~\cite{Koch:2008,Lahaye:2009}. In
the present investigation, we use the parameters of $^{52}$Cr.

It is convenient to rescale Eq.~(\ref{dgpe1}) into a dimensionless form.
For this purpose, we apply the transformation ${\bar{\mathbf{r}}}=\mathbf{r}/l,{\bar{\mathbf{R}}}=\mathbf{R}/l,\bar{a}(t)=a(t)/l,\bar{a}_{\mathrm{dd}}=a_{\mathrm{dd}}/l,\bar{t}=t\bar{\omega}$, $\bar{x}=x/l,\bar{y}=y/l,\bar{z}=z/l,\bar{\phi}=l^{3/2}\phi $, where the harmonic-oscillator length is $l=\sqrt{\hbar /(m\bar{\omega})}$and rewrite Eq. (\ref{dgpe1}) (dropping the
overbar) as
\begin{equation}
\begin{split}
i\frac{\partial \phi ({\mathbf{r}},{t})}{\partial t}=& \biggr[-\frac{1}{2}\nabla ^{2}+V({\mathbf{r}})+4\pi a(t)N|{\phi ({\mathbf{r}},{t})}|^{2}+3Na_{\mathrm{dd}}\int \frac{1-3\cos ^{2}\theta }{|\mathbf{{R}|^{3}}} |\phi ({\mathbf{r}}^{\prime},t)|^2 d{\mathbf{r}}^{\prime } \\& +\chi(t)|{\phi ({\mathbf{r}},{t})}|^{4}\biggr]\phi ({\mathbf{r}},{t}),
\end{split}\label{dgpe2}
\end{equation}
where $\chi(t)=\frac{\eta(t)N^2}{m\bar{\omega}^2l^6}$, 
\begin{equation}
V({\mathbf{r}})=\frac{1}{2}(\gamma^{2}x^{2}+\nu^{2}y^{2}+\lambda^{2}z^{2}),  
\label{Vr}
\end{equation}%
and $\gamma =\omega _{x}/\bar{\omega},\nu =\omega _{y}/\bar{\omega},\lambda =\omega _{z}/\bar{\omega}$, where $\bar{\omega}=(\omega _{x}\omega _{y}\omega _{z})^{1/3}$.

For an axially-symmetric (with $\nu =\gamma $ in Eq. (\ref{Vr})) disk-shaped
DBEC with a strong axial trap ($\lambda >\nu $, $\gamma $), we presume that
the structure of DBEC in the axial direction amounts to the axial ground
state, $\phi _{\mathrm{axial}}(z)=\exp (-z^{2}/2d_{z}^{2})/(\pi
d_{z}^{2})^{1/4},\quad d_{z}\equiv \sqrt{1/\lambda }.$ Hence, the 3D wave function is defined as
\begin{equation}
\phi (\mathbf{r})\equiv \phi _{\mathrm{axial}}(z)\times \psi (\boldsymbol{\rho },t)\equiv (\pi d_{z}^{2})^{-1/4}\exp \left[-z^{2}/\left(2d_{z}^{2}\right) \right] \psi (\boldsymbol{\rho },t), 
\label{phi}
\end{equation}%
where $\boldsymbol{\rho }\equiv (x,y)$, $\psi (\boldsymbol{\rho },t)$ is the
effective 2D wave function and $d_{z}$ is the axial
harmonic-oscillator length. To obtain the effective 2D equation for the
disk-shaped DBEC, we substitute expression (\ref{phi}) in Eq. (\ref{dgpe2}),
multiply it by $\phi _{\mathrm{axial}}(z)$ and integrate the result over $z$
to  arrive at the 2D equation \cite%
{Muruganandam:2012,Lahaye:2008,RKK:2015}

\begin{equation}
\begin{split}
i\frac{\partial \psi (\boldsymbol{\rho },t)}{\partial t}&=\Big[-\frac{1}{2}\nabla _{\rho }^{2}+d(t)V(\mathbf{\rho })+\frac{4\pi a(t)N}{\sqrt{2\pi }d_{z}}|\psi (\boldsymbol{\rho },t)|^{2}+\frac{\chi(t)}{\sqrt{3}\pi d_{z}^2}|\psi (\boldsymbol{\rho },t)|^{4} \\ & +\frac{4\pi a_{\mathrm{dd}}N}{\sqrt{2\pi }d_{z}}\int \frac{d\mathbf{k}_{\rho }}{(2\pi )^{2}}e^{-i\mathbf{k}_{\rho }.\boldsymbol{\rho }}n(\mathbf{k}_{\rho },t)h_{2D}(\sigma )\Big]\psi (\boldsymbol{\rho },t),
\end{split}\label{dgpe3}
\end{equation}
where we have introduced the time-dependent factor $d(t),$ which may
provide temporal modulation of the effective 2D trapping potential  defined as
\begin{equation}
V(\mathbf{\rho })=\frac{\gamma ^{2}x^{2}+\nu ^{2}y^{2}}{2},  \label{Vrho}
\end{equation}

Next, the external trap may be removed adiabatically. For that purpose, the parameter $d(t)$ is slowly ramped down from $1$ to $0$. The dipolar term is written in the Fourier space after calculating the convolution of the corresponding variables~\cite{Muruganandam:2012}, $n(\mathbf{k}_{\rho
},t)=\int d\boldsymbol{\rho }e^{i\mathbf{k}_{\rho }.\boldsymbol{\rho }}|\psi
(\boldsymbol{\rho },t)|^{2}$, $h_{2D}=2-3\sqrt{\pi }\sigma \exp (\sigma
^{2})\{1-\text{erf}(\sigma )\}$, $\sigma =\frac{\mathbf{k}_{\rho }d_{z}}{%
\sqrt{2}}$, $\lambda =9$ and $\mathbf{k}_{\rho }=\sqrt{k_{x}^{2}+k_{y}^{2}}$%
. In Eq.~(\ref{dgpe3}), the length is measured in units of the
characteristic harmonic oscillator length $l=\sqrt{\hbar /m\bar{\omega} }$, time $t$ in units of $\bar{\omega}
^{-1}$ and energy in units of $\hbar \bar{\omega} $. 

The quasi-2D and full 3D nonlocal GP equations are then simulated using the split-step Fourier method. In our dimensionless units, we use the space and time step sizes $\mathbf{0.05}$ and $0.001$. %

\section{The variational method}

\label{sec3} To gain an analytical insight into the condensate dynamics, we
use the variational approximation (VA) with the Gaussian ansatz as a trial
wave function for the solution of Eq.(\ref{dgpe3}) in the absence of an external trap~\cite{Muruganandam:2012}: 

\begin{equation}
\psi (\boldsymbol{\rho },t)=\frac{1}{R(t)\sqrt{\pi }}\exp {\left( -\frac{\rho ^{2}}{2R(t)^{2}}+i\beta(t)\rho ^{2}\right) }, 
\label{ansatz}
\end{equation}%
Here the variables $R(t)$ and $\beta(t)$ are real-valued and stand for the radius and radial chirp of the self-trapped state while the amplitude is determined by the normalization condition  (\ref{N}).
The Lagrangian density generating Eq.~(\ref{dgpe3}) 
in the absence of the trap ($d(t)=0$) is given by

\begin{eqnarray}
\mathcal{L}=&\displaystyle\frac{i}{2}\left( \psi \frac{\partial \psi ^{\ast}}{\partial t}-\psi ^{\ast }\frac{\partial \psi }{\partial t}\right) +\frac{|\nabla _{\rho }\psi |^{2}}{2}+\frac{2\pi a(t)N}{\sqrt{2\pi }d_{z}}|\psi |^{4}+\frac{\chi(t)}{3\sqrt{3}\pi d_{z}^2}|\psi |^{6}  \notag \\ &\,\displaystyle+\frac{2\pi a_{dd}N}{\sqrt{2\pi }d_{z}}|\psi |^{2}\int \frac{d\mathbf{k}_{\rho }}{(2\pi )^{2}}e^{i\mathbf{k}_{\rho }.\boldsymbol{\rho }}n(\mathbf{k}_{\rho},t)h_{2D}(\sigma )\Bigg). 
\label{Lagd}
\end{eqnarray}
where $\psi^{\ast}$ denotes the complex conjugate of the wave function $\psi$. The trial wave function (\ref{ansatz}) is substituted into the  Lagrangian density (\ref{Lagd}) and the effective Lagrangian is calculated by integrating the Lagrangian density. Then, the following Euler-Lagrangian equations for the variational parameters can be derived from the effective Lagrangian as 
\begin{align}
\frac{dR(t)}{dt}&=2R(t)\beta(t), \label{rt}\\ 
\frac{d\beta(t)}{dt}&= \frac{1}{2R(t)^{4}}+\frac{Na(t)}{\sqrt{2\pi}d_{z}R(t)^{4}}-\frac{Na_{dd}\Lambda (\xi(t) )}{2\sqrt{2\pi }d_{z}R(t)^{4}}  +\frac{2N^{2}\chi (t)}{9\sqrt{3}d_{z}^{2}\pi ^{3}R(t)^{6}}-2\beta(t)^2. 
\label{bt}
\end{align}
By associating Eqs. (\ref{rt}) and (\ref{bt}), we obtain the following equation for the evolution of the width $R(t)$ as
\begin{equation}
\begin{split}
\frac{d^{2}R(t)}{dt^{2}}=& \frac{1}{R(t)^{3}}+\frac{\sqrt{2}Na(t)}{\sqrt{\pi}d_{z}R(t)^{3}}-\frac{Na_{dd}\Lambda (\xi(t) )}{\sqrt{2\pi }d_{z}R(t)^{3}}  +\frac{4N^{2}\chi (t)}{9\sqrt{3}d_{z}^{2}\pi ^{3}R(t)^{5}}
\end{split}%
\end{equation}%
where $\Lambda (\xi(t) )=2-7\xi(t)^{2}-4\xi(t)^{4}+9\xi(t)^{4}d(\xi(t))/(1-\xi(t)^{2})^{2} $, $\xi(t)=R(t)/d_z$. 

We plan to consider the effects of periodic modulation of the contact interaction in the form of
\begin{equation}
a(t)=\epsilon _{0}+\epsilon _{1}\sin {(\Omega t)},\,\,\chi (t)=\chi _{0}+\chi_{1}\sin {(\Omega t)}  
\label{sin}
\end{equation}
on the stability of DBEC where $\epsilon _{0}$, $\chi _{0}$ are constant parts of the two-body and three-body interactions respectively and $\epsilon _{1}$, $\chi _{1}$ are the amplitudes of their temporal modulation. The Kapitza averaging method can be employed to treat the oscillatory parts~\cite{Landau:1960}.
With the explicitly included oscillating nonlinearity, we get the following equation for the evolution of the radial width (cf. Ref. \cite%
{Muruganandam:2012}):

\begin{equation}
\begin{split}
\frac{d^{2}R(t)}{dt^{2}}& =\frac{1}{R(t)^{3}}+\frac{\sqrt{2}N[\epsilon_{0}+\epsilon _{1}\sin {(\Omega t)}]}{\sqrt{\pi }d_{z}R(t)^{3}}-\frac{Na_{dd}\Lambda (\xi(t) )}{\sqrt{2\pi }d_{z}R(t)^{3}}  +\frac{4N^{2}[\chi _{0}+\chi _{1}\sin {(\Omega t)}]}{9\sqrt{3}d_{z}^{2}\pi^{3}R(t)^{5}}
\end{split}%
\end{equation}%
Now, $R(t)$ may be separated into a slowly varying part $R_{0}(t)$ and a rapidly varying one $R_{1}(t),R(t)=R_{0}(t)+R_{1}(t)$. When $\Omega \ \gg 1$, $R_{1}(t)$ is small of the order of $\sim $ $\Omega ^{-2}$. 
Keeping the terms of the order of  $\Omega^{-2}$ in $R_1(t)$, one may obtain the following equations of motion for $R_0(t)$ and $R_1(t)$,

\begin{align}
\frac{d^{2}R_1(t)}{dt^{2}} =&\frac{\sqrt{2}N\epsilon _{1}\sin {(\Omega t)}}{\sqrt{\pi }d_{z}R_0^{3}}+\frac{4N^{2}\chi _{1}\sin {(\Omega t)}}{9\sqrt{3}d_{z}^{2}\pi^{3}R_0^{5}} \label{r1dd} \\
\frac{d^{2}R_{0}(t)}{dt^{2}}=& \frac{1}{R_{0}^{3}}+\frac{N[2\epsilon_{0}-a_{dd}\,\Lambda (\xi_{0}(t))]}{\sqrt{2\pi }d_{z}\,R_{0}^{3}}+\frac{4N^{2}\chi _{0}}{9\sqrt{3}\pi ^{3}d_{z}^{2}\,R_{0}^{5}}- \frac{3\sqrt{2}N\epsilon_{1}<R_1\,sin(\Omega t)>}{\sqrt{\pi} d_{z}\,R_{0}^{4}} \label{r0dd}\\ \nonumber
& -\frac{20N^{2}\chi _{1}<R_1\,sin(\Omega t)>}{9\sqrt{3}\pi^{3}d_{z}^{2}\,R_{0}^{6}}. 
\end{align}%
where the $<R_1\,sin(\Omega t)>$ indicates the time average of the rapid oscillation. From Eq.(\ref{r1dd}), we obtain

\begin{equation}
R_1(t) =-\biggr[\frac{\sqrt{2}N\epsilon _{1}\sin {(\Omega t)}}{\sqrt{\pi }d_{z}\Omega^2R_0^{3}}+\frac{4N^{2}\chi _{1}\sin {(\Omega t)}}{9\sqrt{3}d_{z}^{2}\pi^{3}\Omega^2R_0^{5}}\biggr].  
\label{r1}
\end{equation}%
We now  substitute $R_1(t)$ into Eq.(\ref{r0dd}) to  obtain the following equation of motion for $R_0(t)$
\begin{equation}
\begin{split}
\frac{d^{2}R_{0}(t)}{dt^{2}}=& \frac{1}{R_{0}(t)^{3}}+\frac{N[2\epsilon_{0}-a_{dd}\,\Lambda (\xi_{0}(t))]}{\sqrt{2\pi }d_{z}\,R_{0}(t)^{3}}+\frac{4N^{2}\chi _{0}}{9\sqrt{3}\pi^{3}d_{z}^{2}\,R_{0}(t)^{5}}+ \frac{3N^{2}\epsilon_{1}^{2}}{\pi d_{z}^{2}\,\Omega ^{2}\,R_{0}(t)^{7}}\\ & +\frac{16\sqrt{2}N^{3}\epsilon _{1}\chi_{1}}{9\sqrt{3\pi^{7}}d_{z}^{3}\,\Omega ^{2}\,R_{0}(t)^{9}}+\frac{40N^{4}\chi _{1}^{2}}{243\pi^{6}d_{z}^{4}\,\Omega ^{2}\,R_{0}(t)^{11}}.
\end{split}%
\end{equation}%
The VA suggests that the effect of the DD interaction is to minimize the impact of the constant part of the contact interaction for $a_{dd}>0$. Thus, one can conclude that the system may becomes effectively attractive depending on the choice of $a_{dd}$ and $\epsilon _{0}$. Hence one can explore the possibility of creation of bright solitons even for positive (repulsive) scattering lengths provided  $a_{dd}>\epsilon _{0}$. 
Thus, it is pretty obvious that DD interaction minimizes the effect of contact interaction as dipolar BECs really need non-zero dipolar strength to compete with contact interaction. In addition,  DD interaction ($a_dd>0$) always counteracts with two-body contact interaction.
 
We do agree under   the present circumstances that  the dipoles assume   a pancake that are  
oriented in the z-direction and dipolar BEC expands along the xy plane. To arrest the expansion, here we use the attractive two-body contact interaction. The effective potential $U(R_{0})$ corresponding to the above
equation of motion can be written as
\begin{equation}
\begin{split}
U(R_{0})=& \frac{1}{2R_{0}^{2}}+\frac{N[2\epsilon _{0}-a_{dd}\,\delta (\xi_{0})]}{2\sqrt{2\pi }d_{z}\,R_{0}^{2}}+\frac{N^{2}\chi _{0}}{9\sqrt{3}\pi^{3}d_{z}^{2}\,R_{0}^{4}}+ \frac{N^{2}\epsilon _{1}^{2}}{2\pi d_{z}^{2}\,\Omega ^{2}\,R_{0}^{6}}+\frac{2\sqrt{2}N^{3}\epsilon _{1}\chi _{1}}{9\sqrt{3\pi ^{7}}d_{z}^{3}\,\Omega ^{2}\,R_{0}^{8}}\\ & +\frac{4N^{4}\chi _{1}^{2}}{243\pi^{6}d_{z}^{4}\,\Omega ^{2}\,R_{0}^{10}}. 
\end{split} \label{effpt}
\end{equation}

\noindent where $\delta (\xi _{0})=[1+2\xi _{0}^{2}-3\xi _{0}^{2}d(\xi
_{0})]/(1-\xi _{0}^{2})$ and $\xi _{0}=R_{0}/d_{z}$. Now, one can scrutinize the nature of the effective potential for different system parameters, namely the strengths of the two-body, three-body and DD interactions. In the present analysis, we explore the possibility of stabilizing the trapless dipolar condensates by manipulating the constant and oscillatory part of the two- and three-body interactions.
\begin{figure}[t]
\begin{center}
\includegraphics[width=6.6cm,height=4.8cm]{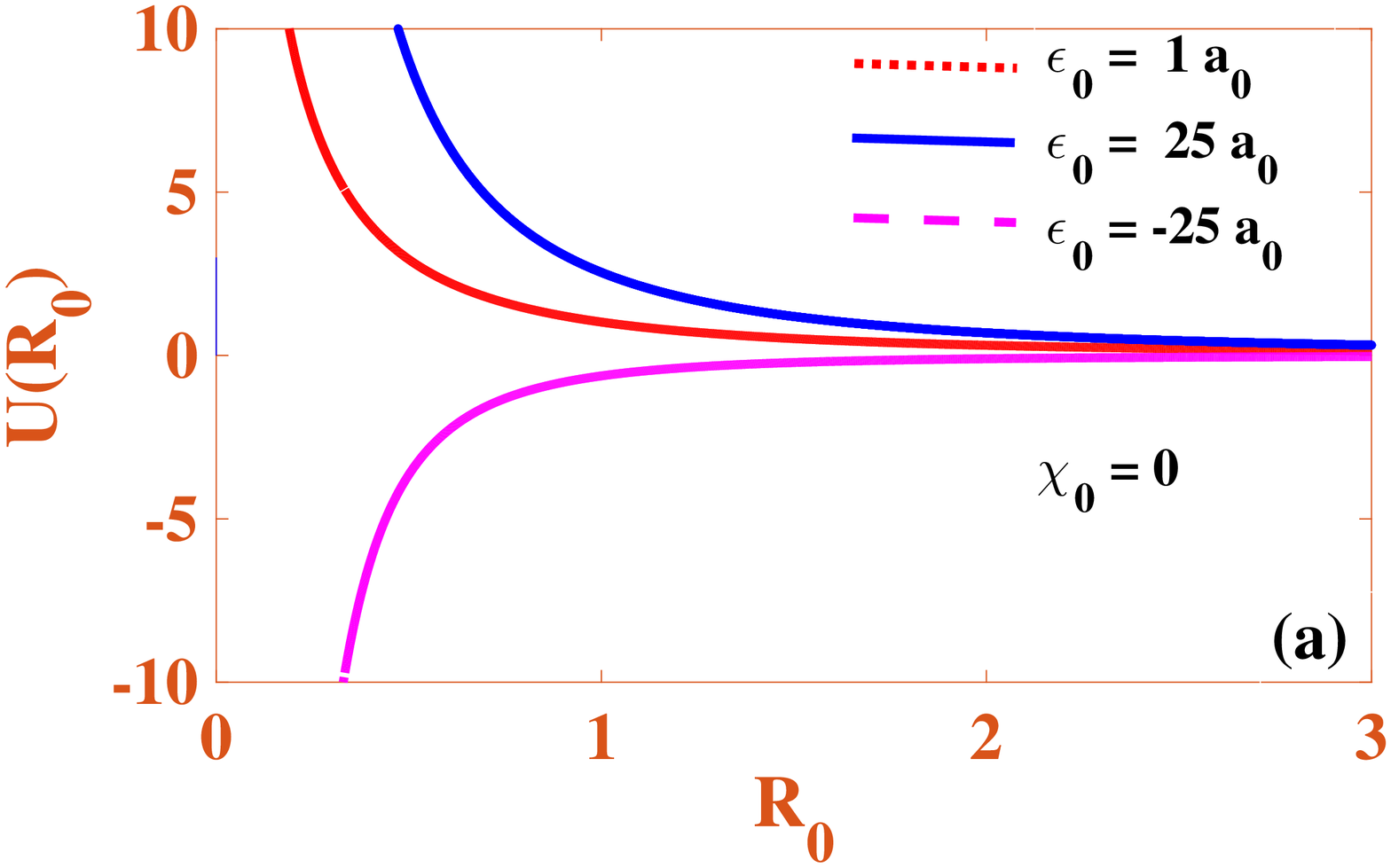}
\includegraphics[width=6.6cm,height=4.8cm]{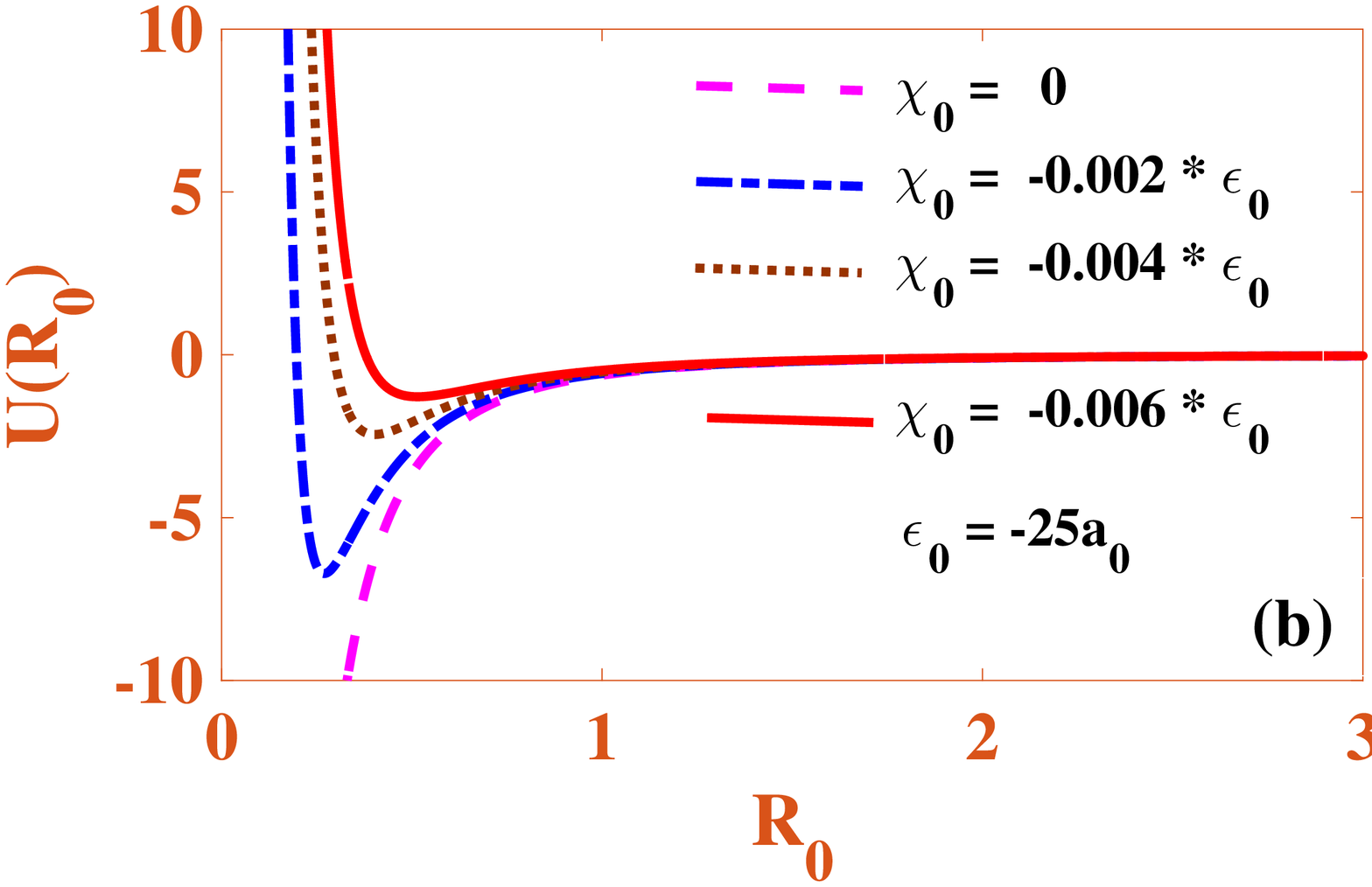}
\end{center}
\caption{ (a) Potential curves for $\protect\epsilon _{0}=$ 1$a_{0}$, 25$%
a_{0}$, -25$a_{0}$ and the other parameters are $\protect\epsilon _{1}=0$, $%
\protect\chi _{0}$, $\protect\chi _{1}=0$ (b) Potential curves for $\protect%
\chi _{0}=0$, -0.002$\protect\epsilon _{0}$, -0.004$\protect\epsilon _{0}$
and -0.006$\protect\epsilon _{0}$ along with $\protect\epsilon _{1}=0$ and $%
\protect\chi _{1}=0$. }
\label{f1}
\end{figure}

\begin{figure}[hb!]
\begin{center}
\includegraphics[width=0.85\linewidth]{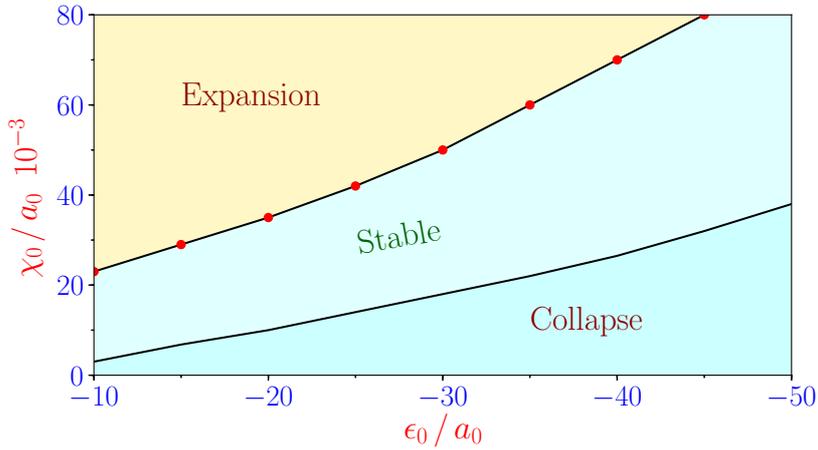}
\end{center}
\caption{Stability domains for constant part of the two-body interaction vs constant part of the three-body interaction. The remaining parameters are $\protect\epsilon _{1}=0$ and $\protect\chi _{1}=0$.}
\label{f2}
\end{figure}
\begin{figure}[ht!]
\begin{center}
\includegraphics[width=0.8\linewidth]{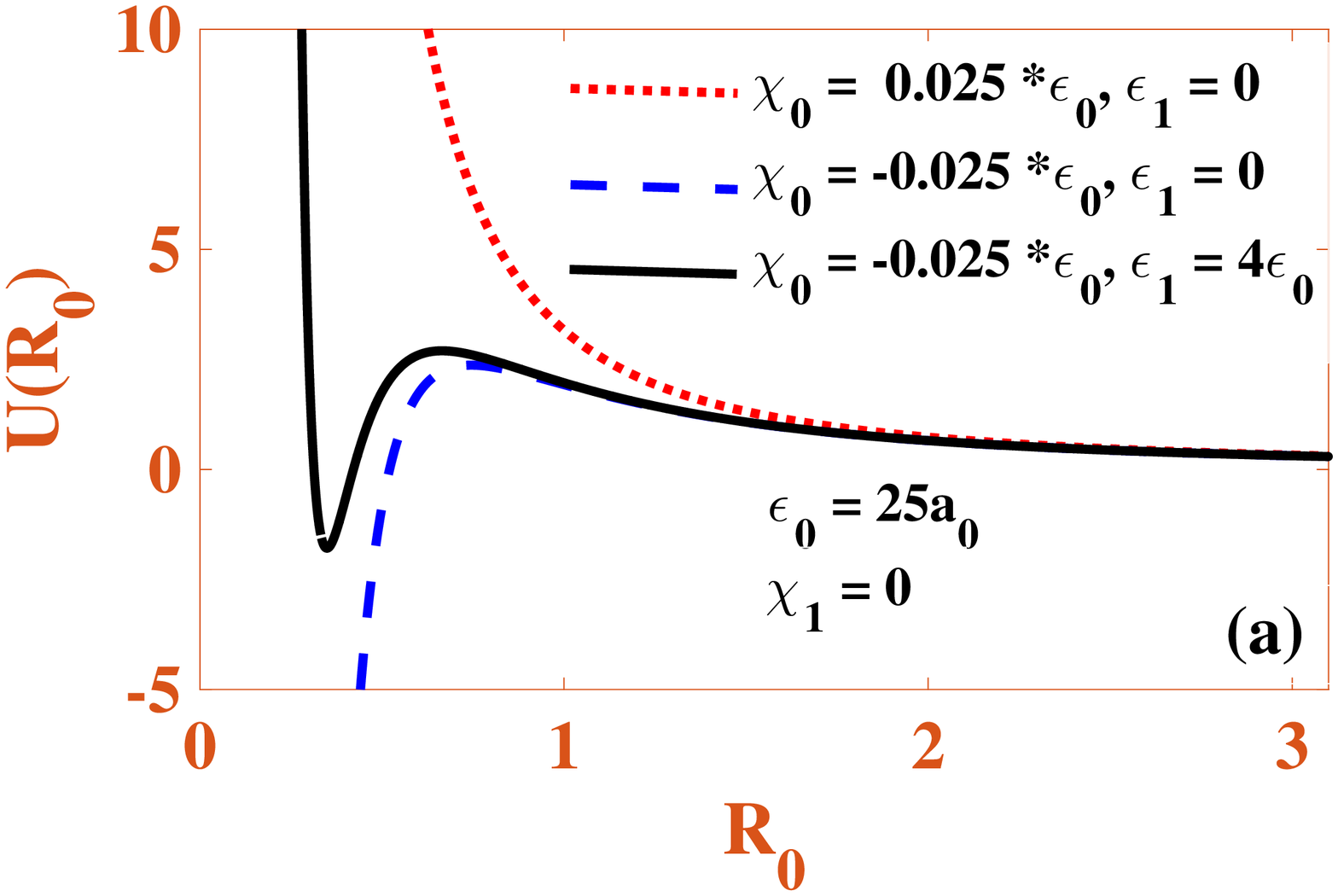} 
\includegraphics[width=0.8\linewidth]{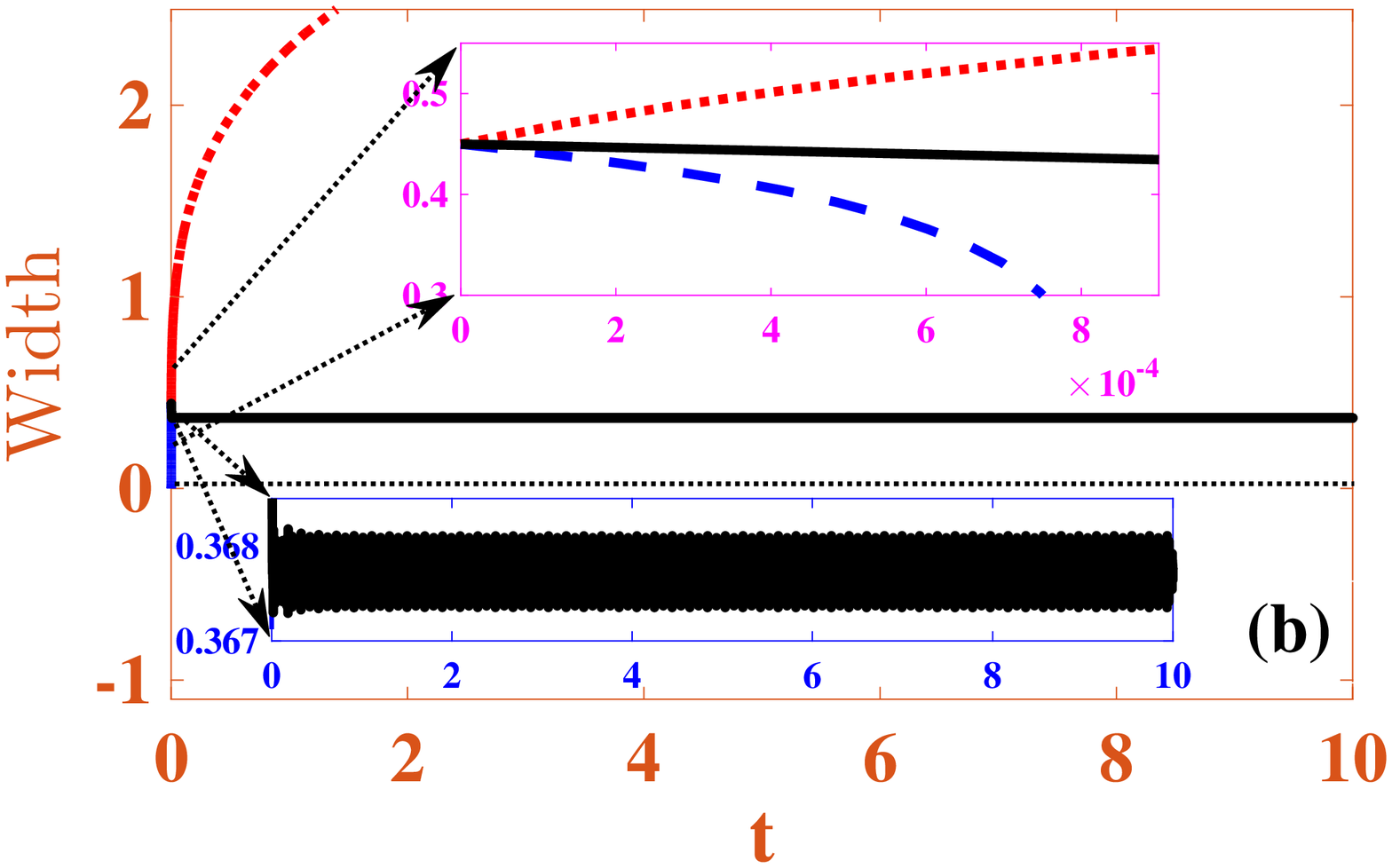}
\end{center}
\caption{(a) Plot of the effective potential U($R_0$) versus $R_0$ and (b) the corresponding time dependence of the slowly varying width $R_0$. }
\label{f3}
\end{figure}

\begin{figure}[hb!]
\begin{center}
\includegraphics[width=0.8\linewidth]{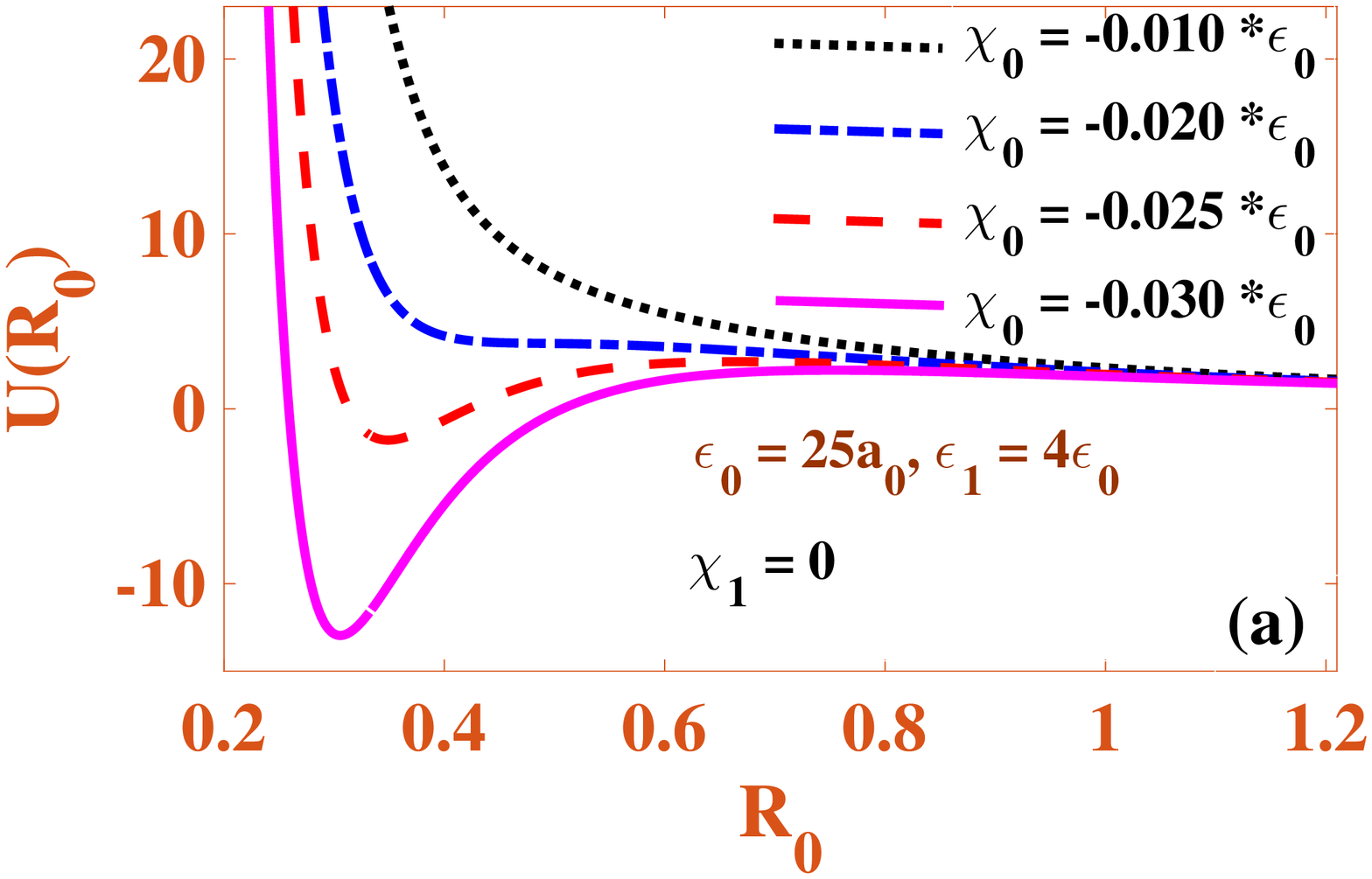} %
\includegraphics[width=0.8\linewidth]{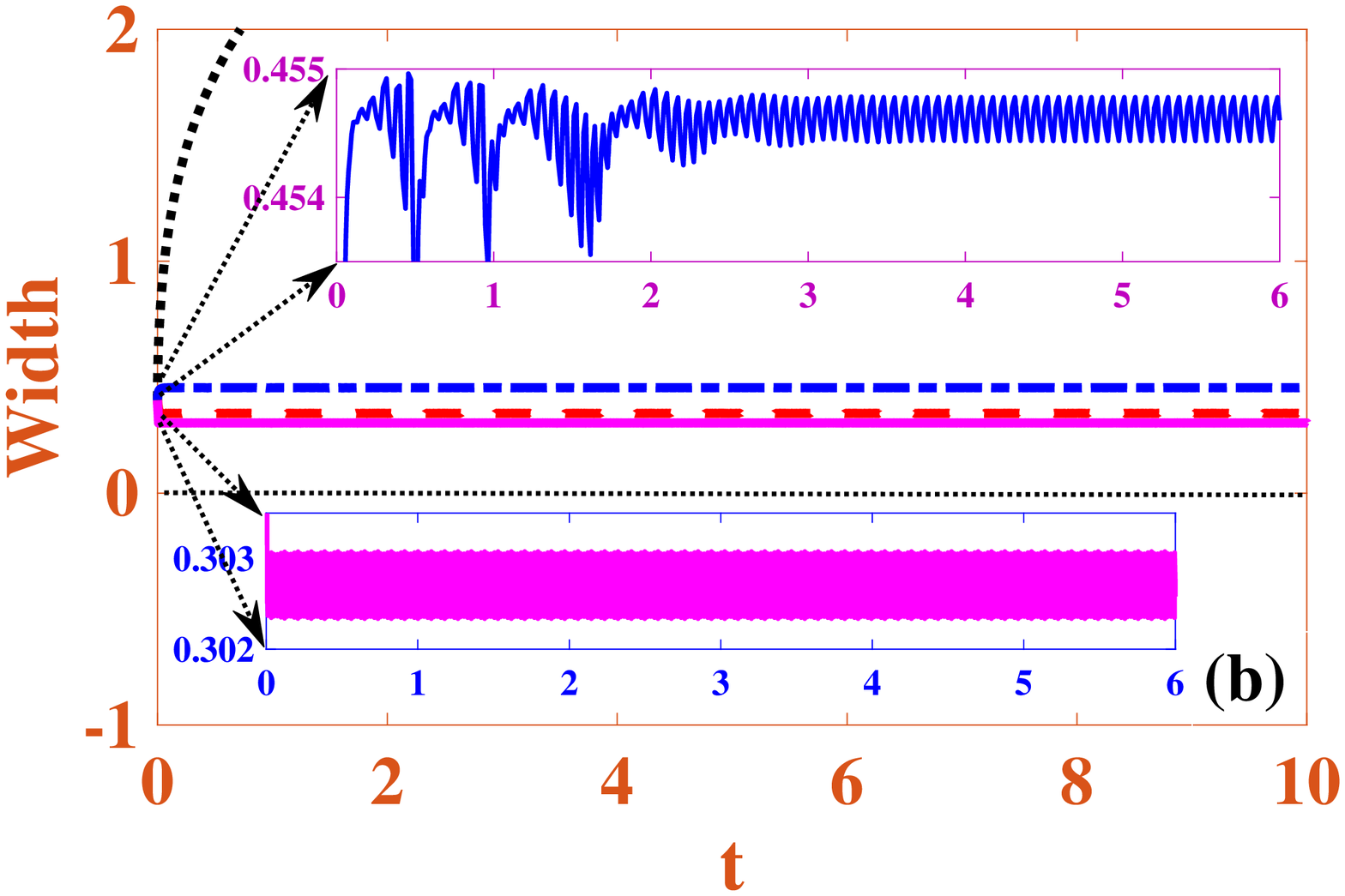}
\end{center}
\caption{(a) Plot of the effective potential U($R_0$) versus $R_0$ and (b) The corresponding time dependence of the slowly varying width $R_0$. }
\label{f4}
\end{figure}
\begin{figure}[hb!]
\begin{center}
\includegraphics[width=0.8\linewidth]{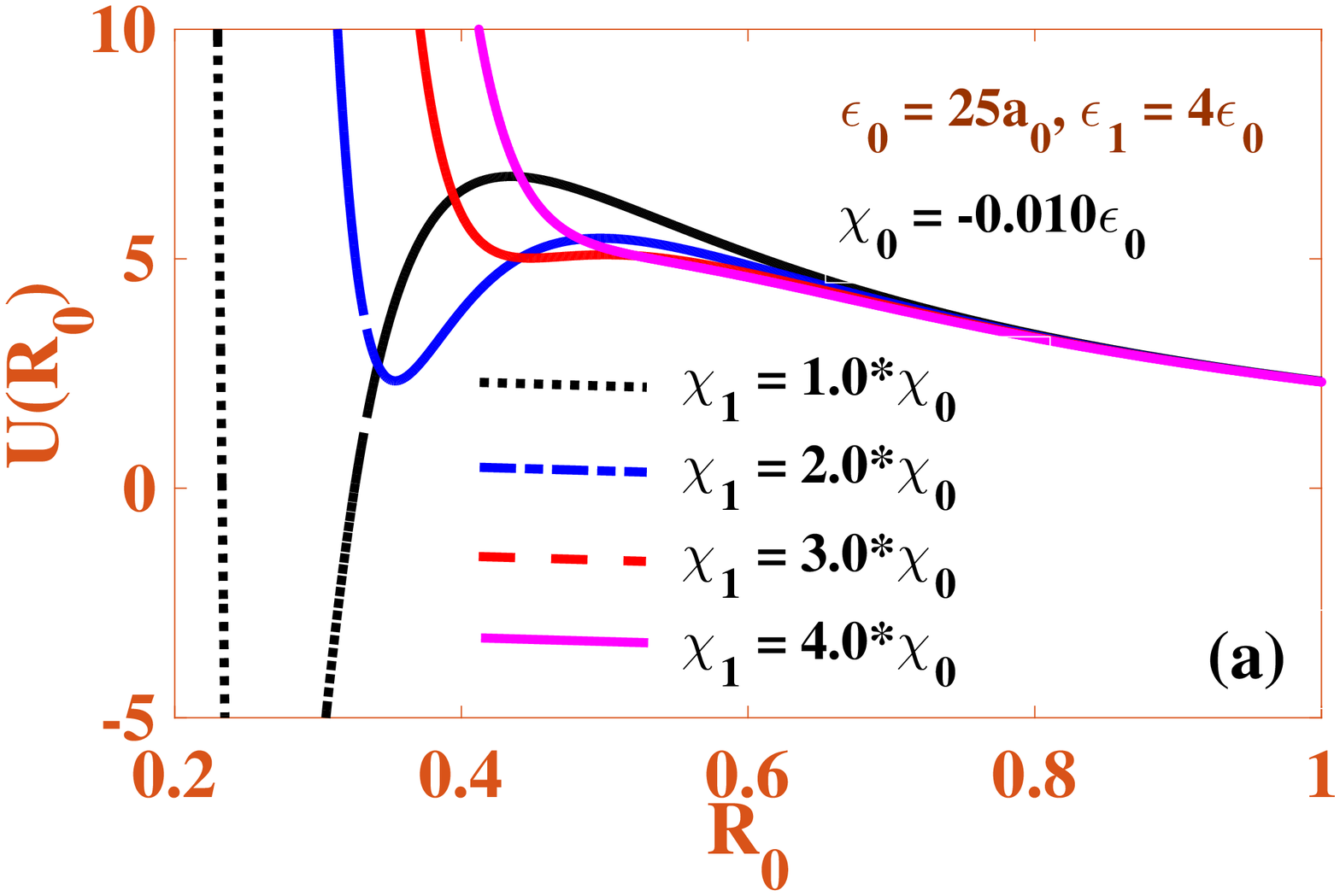}
\includegraphics[width=0.8\linewidth]{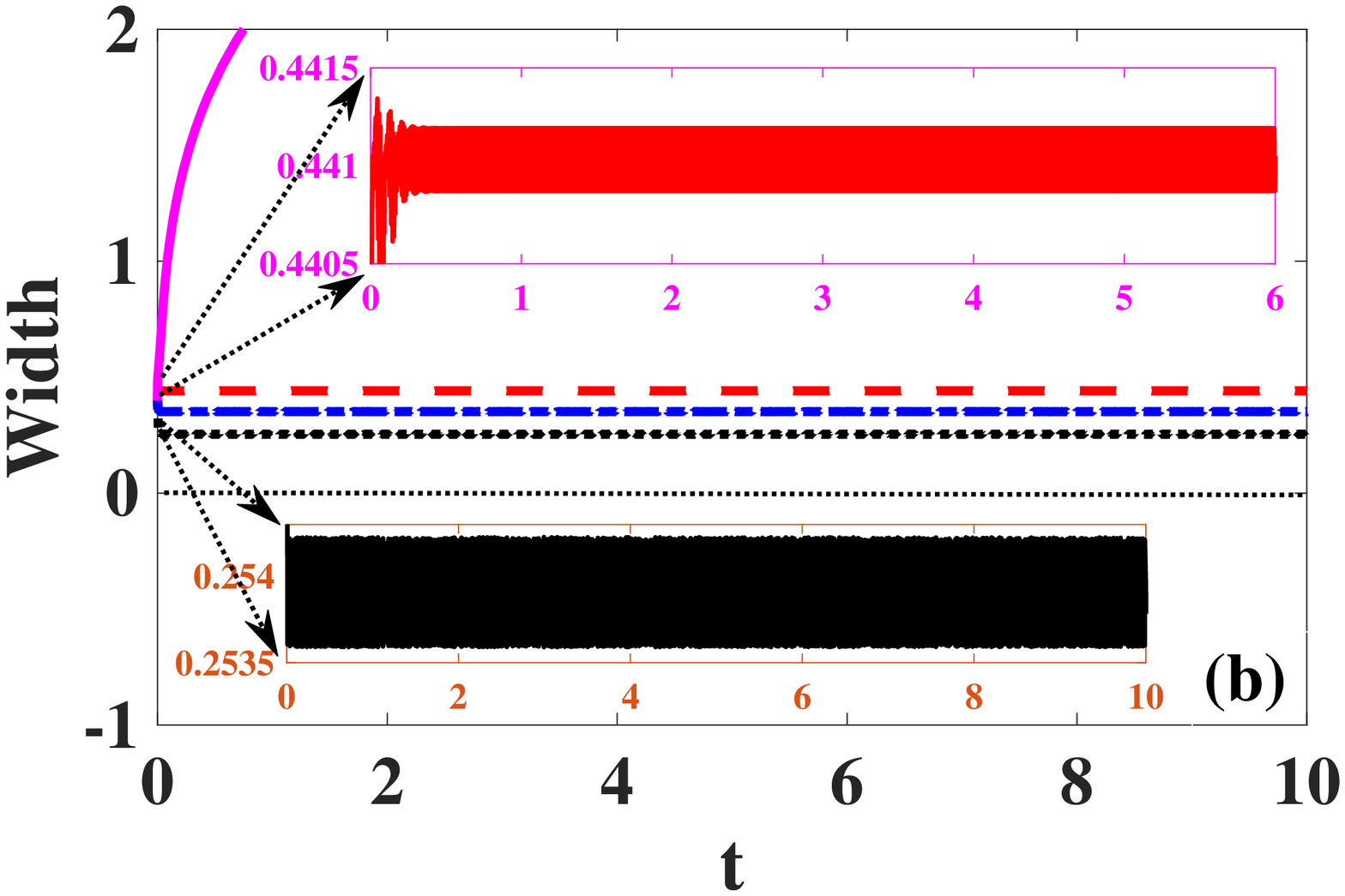}
\end{center}
\caption{(a) Plot of the effective potential U($R_0$) versus $R_0$ for $^{52}$Cr. (b) The equilibrium width $R_0$ as a function of time. }
\label{f5}
\end{figure}

In Fig.~\ref{f1}, we display the effective potential (\ref{effpt}) of  the $^{52}$Cr condensate in the absence of the temporal modulation of both nonlinearities. One may infer from Fig.~\ref{f1}(a) that the potential energy curves do not show any minimum for both repulsive (dotted and solid curves) and attractive (the dashed curvraction ($\epsilon_{0}$) strengths. In the repulsive case, this indicates that the condensate leads to expansion due to the repulsive two-body interaction and the kinetic pressure. On the other hand, no stable counter balance point (i.e., potential minimum) exists for the attractive case either, which indicates at the attractive constant two-body interaction overcomes the kinetic pressure, leading to the collapse of the condensates. These conclusions are commonly known for the cubic GP/nonlinear-Schr\"{o}dinger equation in 2D \cite{Fibich}%

Next, in Fig.~\ref{f1}(b), we show how the interplay between the two-and three-body collisions along with the DD interaction can lead to a stable condensate. We first get back to one of the potential curves in Fig.~\ref{f1}(a) for the attractive case corresponding to $\epsilon _{0}=-25a_{0}$, $\epsilon _{1}=0$, $\chi _{0}=0$, $\chi _{1}=0$. In this case, the absence of a repulsive force to balance the self-attraction force arising due to the constant part of the two-body and DD interactions initiates the collapse of the condensates. Now, the combined impact of the repulsive force arising due to the repulsive three-body interaction safely counterbalances the self-attraction as is evidenced by the presence of the potential minima in the respective curve in Fig.~\ref{f1}(b) while the nonlinearity management is not included.
Further, in Fig~\ref{f2}, we show the stability  domains in the plane of the constant parts of the {\it attractive two-body interaction} in competition with {\it repulsive three-body interactions}. The diagrams show stable regimes between the collapse and expansion regions. From figs.~\ref{f1}-\ref{f2}, we conclude that with help of the repulsive three-body interaction, one can stabilize the trapless DBEC with attractive two-body interaction. This has been done without the contribution of the oscillatory part of the contact interactions. 

Next, we plan to stabilize the trapless DBECs with repulsive two-body contact interaction. We first think about one of the potential energy curves from Fig.~\ref{f1}(a) for the repulsive case corresponding to $\epsilon _{0}=+25a_{0}$, $\epsilon _{1}=0$, $\chi _{0}=0$, $\chi _{1}=0$. Under these circumstances, the additional repulsive force caused by the constant part the three-body interaction along with the constant part the two-body interaction and the DD interaction makes the condensate decay through the expansion as shown in Fig.~\ref{f3}(a) by the dotted red curve. Now to stabilize the system in this case, we introduce an attractive force arising from the constant part of the three-body interaction, $\chi_{0}<0$. We observe the presence of an inverted dashed blue-colored potential curve in Fig.~\ref{f3}(a) implying that we have not yet reached a stable regime. Moreover, a stable region has been obtained neither with the help of an attractive nor with that of a repulsive three-body interaction. However, the introduction of the time-dependent part of the two-body interaction along with the attractive three-body interaction recovers potential energy minima represented by the black curve in Fig.~\ref{f3}(a). Here, the two-body repulsion is balanced by the interplay between the attractive three-body interaction and the time-modulated part of the two-body interaction, $\epsilon _{1}$. Figure~\ref{f3}(b) depicts the time evolution of the width of the condensate corresponding to Fig.~\ref{f3}(a). %

Figure~\ref{f3}(b) clearly shows the dynamical regimes corresponding to the expansion, collapse and stability of DBEC. In particular, the stable case with $\chi_{0}=-0.025$ and $\epsilon _{1}=4\epsilon _{0}$ displays robust small-amplitude oscillations of the width. On the other hand, the expansion and collapse are observed, respectively, at $\chi_{0}=0.025$, $\epsilon_{1}=0$, and $\chi_{0}=-0.025$, $\epsilon _{1}=0$. Further, in Figs.~\ref{f4}(a) and (b), we show the effective potential for different values of the constant part of the three-body interaction and the corresponding VA-predicted evolution of the width of the condensate. Once again, robust oscillations of the widths confirm the stability of the trapless DBECs which corresponds to the presence of the minimum in the potential curve.

We further show that the introduction of the time-dependent part of the three-body interaction, $\chi_{1}$ may further enhance the stability of the trapless dipolar repulsive BECs. Figures~\ref{f5} (a) and (b) display the effective potential for various values of $\chi _{1}$ and the evolution of the width of the condensate respectively. Figure~\ref{f5}(a) confirms that one can make the potential well deeper to  stabilize the condensate. In Fig.\ref{f4}(a), the potential minimum starts to appear for $\chi_0=-0.020\epsilon_0$ which is plotted as dash-dotted blue curve. However, no potential minimum exists for $\chi_0=-0.010\epsilon_0$ which is plotted as dotted black curve. From the above, we conclude that the potential depth starts to appear for $\chi_0\geq |-0.020\epsilon_0$|. Also, if we increase $\chi_0$ to $-0.025\epsilon_0$, and $-0.030\epsilon_0$, the potential depth increases further. From this, it is obvious that if the attraction becomes too strong, the system becomes highly unstable leading to its  collapse.

But, the inclusion of the $\chi_1$ along with the case discussed in Fig.\ref{f4}(a) reduces the strong attraction. 
If the attraction is weak for its size, the system becomes weakly attractive in the final stage and it expands to infinity.
On the other hand, if we include a suitable $\chi_1$, then the attraction due to $\chi_0$ is balanced by the oscillation due to the effect of $\chi_1$ and the system becomes stable as shown in Fig.\ref{f5}. 
In Table 1, we bring out the stability domains of trapless DBEC under the combined impact of both constant and oscillatory parts of the two- and three-body interactions.
\color{black}
\begin{table}[h]
\caption{Stability properties of $^{52}$Cr trapless dipolar BECs.}
\label{table2}
\begin{center}
\begin{tabular}{|c|c|c|c|c|c||}
\hline\hline
$\,\,\,\epsilon_0 \,\,\,$ & $\,\,\, \epsilon_1 \,\,\,$ & $\,\,\, \chi_0
\,\,\,$ & $\,\,\, \chi_1 \,\,\,$ & \,\,\,Minimum \, & \,\, Inference \\
\hline\hline
1$a_0$ & 0 & 0 & 0 & No & Unstable \\
&  &  &  &  & (Expansion) \\
25$a_0$ & 0 & 0 & 0 & No & Unstable \\
&  &  &  &  & (Expansion) \\
-25$a_0$ & 0 & 0 & 0 & No & Unstable \\
&  &  &  &  & (Collapse) \\ \hline\hline
$-ve$ & 0 & $+ve$ & 0 & Yes & Stable \\
(-25$a_0$) &  & (-0.002$\epsilon_0$) &  &  &  \\ \hline\hline
$+ve$ & 0 & $-ve$ & 0 & No & Unstable \\
(25$a_0$) &  & (-0.025 $\epsilon_0$) &  &  & (Collapse) \\ \hline
$+ve$ & $\neq 0$ & $-ve$ & 0 & Yes & Stable \\
(25$a_0$) & (4 $\epsilon_0$) & (-0.025 $\epsilon_0$) &  &  &  \\ \hline\hline
$+ve$ & $\neq 0$ & $-ve$ & $\neq 0$ & Yes & Stable \\
(25$a_0$) & (4 $\epsilon_0$) & (-0.01$\epsilon_0$) & (2$\chi_0$) &  &  \\
\hline\hline
\end{tabular}%
\end{center}
\end{table}

\section{Three-dimensional Numerical Results}

\label{sec4}

In this section, we consider  the  full 3D model for direct numerical simulations of Eq.(\ref{gpe3d}). First, we find the ground-state solution by solving Eq. (\ref{gpe3d}) by means of the imaginary-time ($t\rightarrow -\mathrm{i}t$) propagation in the presence of the external trap and two-body contact plus dipolar interactions in the absence of  three-body interactions. Then, the evolution of the input provided by the ground state is simulated in the real time. In the course of the real-time evolution, the external trap is adiabatically removed and the three-body interaction is ramped up which brings the system to the same form for which the VA was elaborated above.
We now consider the following 3D GP equation 

\begin{equation}
\begin{split}
i\frac{\partial \phi ({\mathbf{r}},{t})}{\partial t}&=\biggr[-\frac{1}{2}\nabla^{2}+d(t)\frac{1}{2}(\gamma ^{2}x^{2}+\nu^{2}y^{2}+\lambda^{2}z^{2})+g(t)|{\phi({\mathbf{r}},{t}), }|^{2}+ \\ & \chi (t)|{\phi ({\mathbf{r}},{t}),}|^{4}+3Na_{\mathrm{dd}}\int \frac{1-3\cos ^{2}\theta }{|\mathbf{{R}|^{3}}}|\phi ({\mathbf{r}}^{\prime},t)|^2 d{\mathbf{r}}^{\prime }\biggr]\phi ({\mathbf{r}},{t}). 
\end{split} \label{gpe3d}
\end{equation} 
and numerically solve by  gradually decreasing $d(t)$ and increasing $g(t)$ , $\chi (t)$.  
The DD interaction was evaluated by means of the fast Fourier transform~\cite{Goral:2002}. Typical space and time steps for the numerical grid were taken as $0.05$ and $0.005$, respectively.

\begin{figure}[th]
\begin{center}
\includegraphics[width=0.35\textwidth]{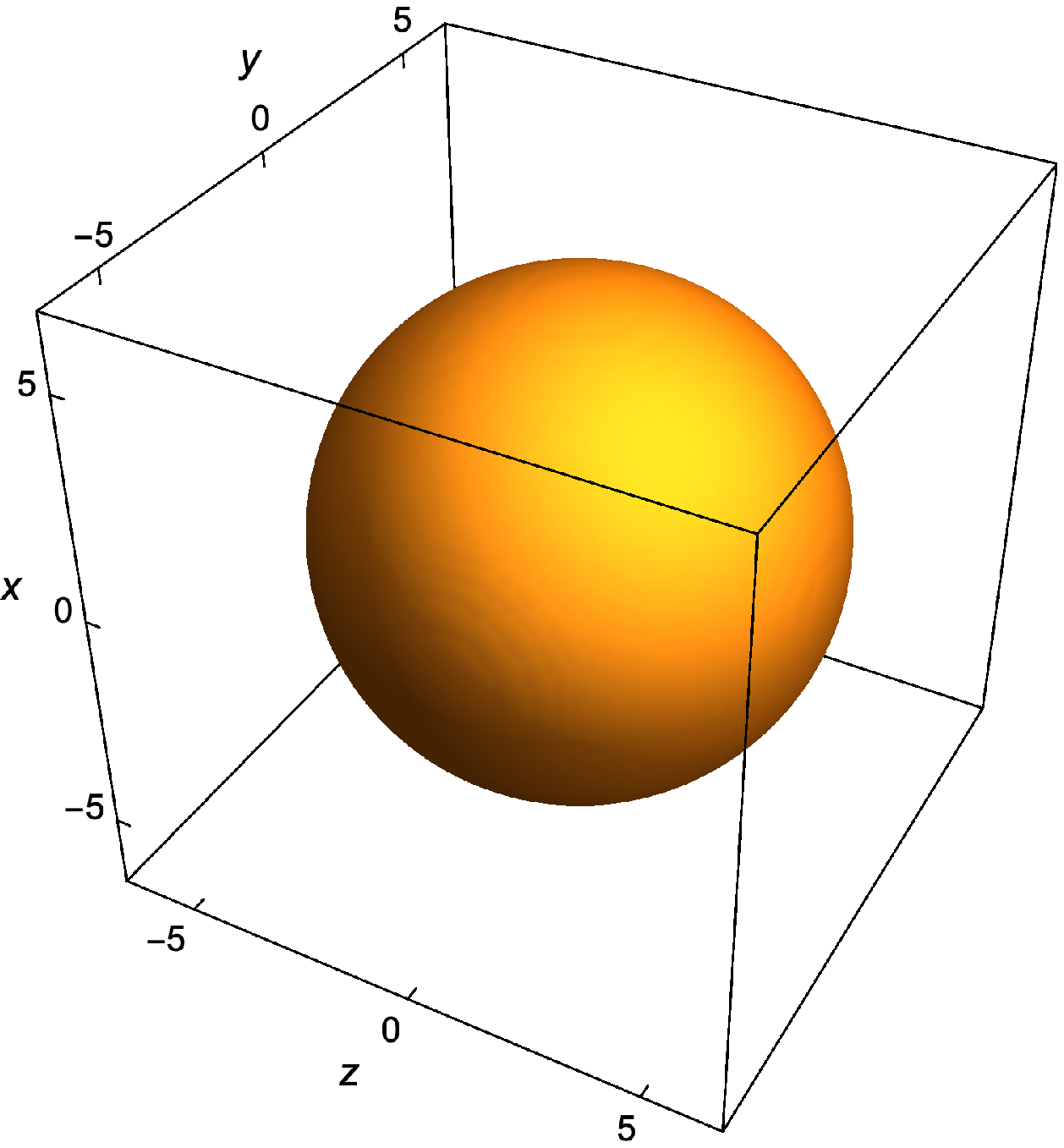} \includegraphics[width=0.35\textwidth]{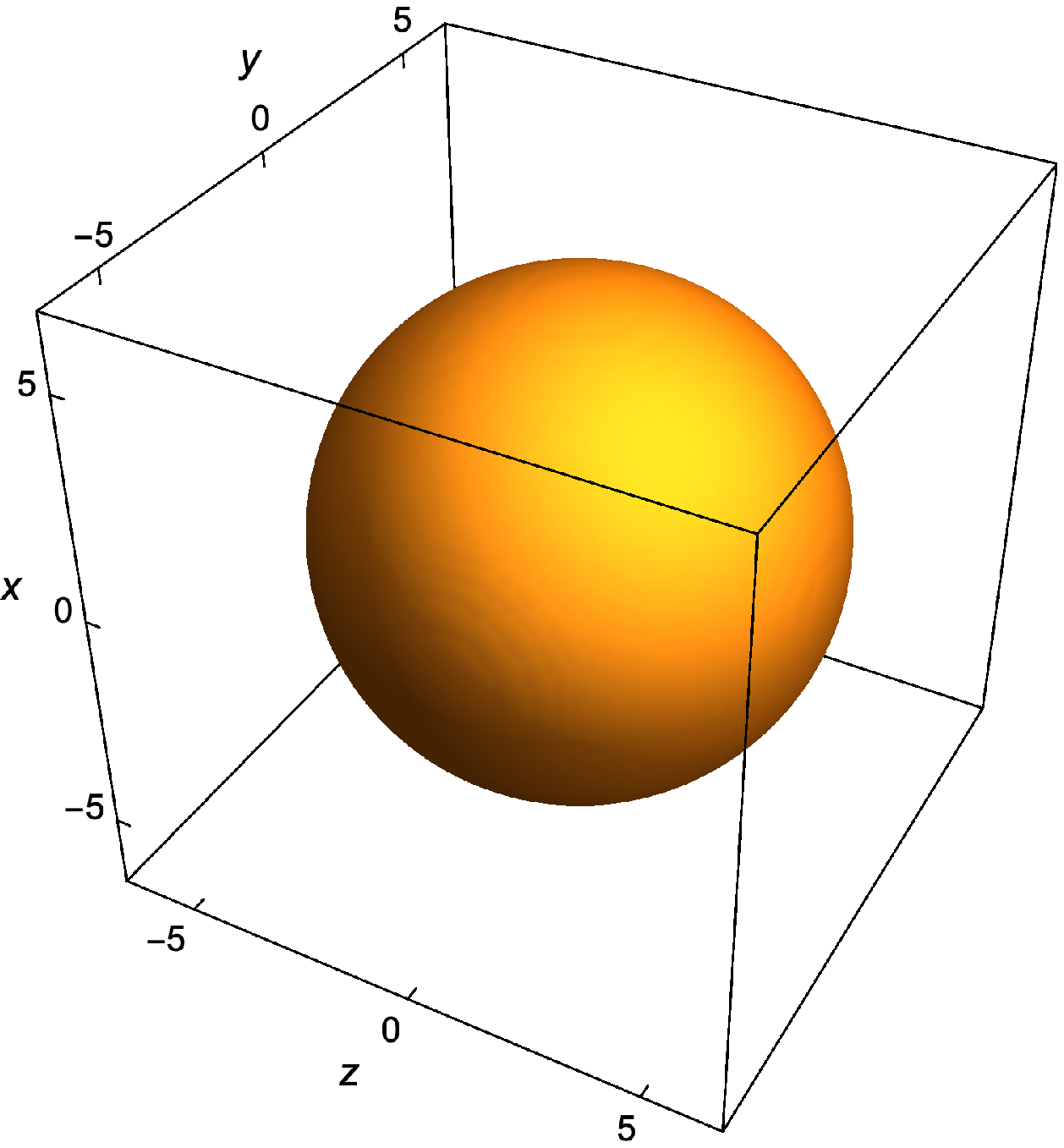} \includegraphics[width=0.35\textwidth]{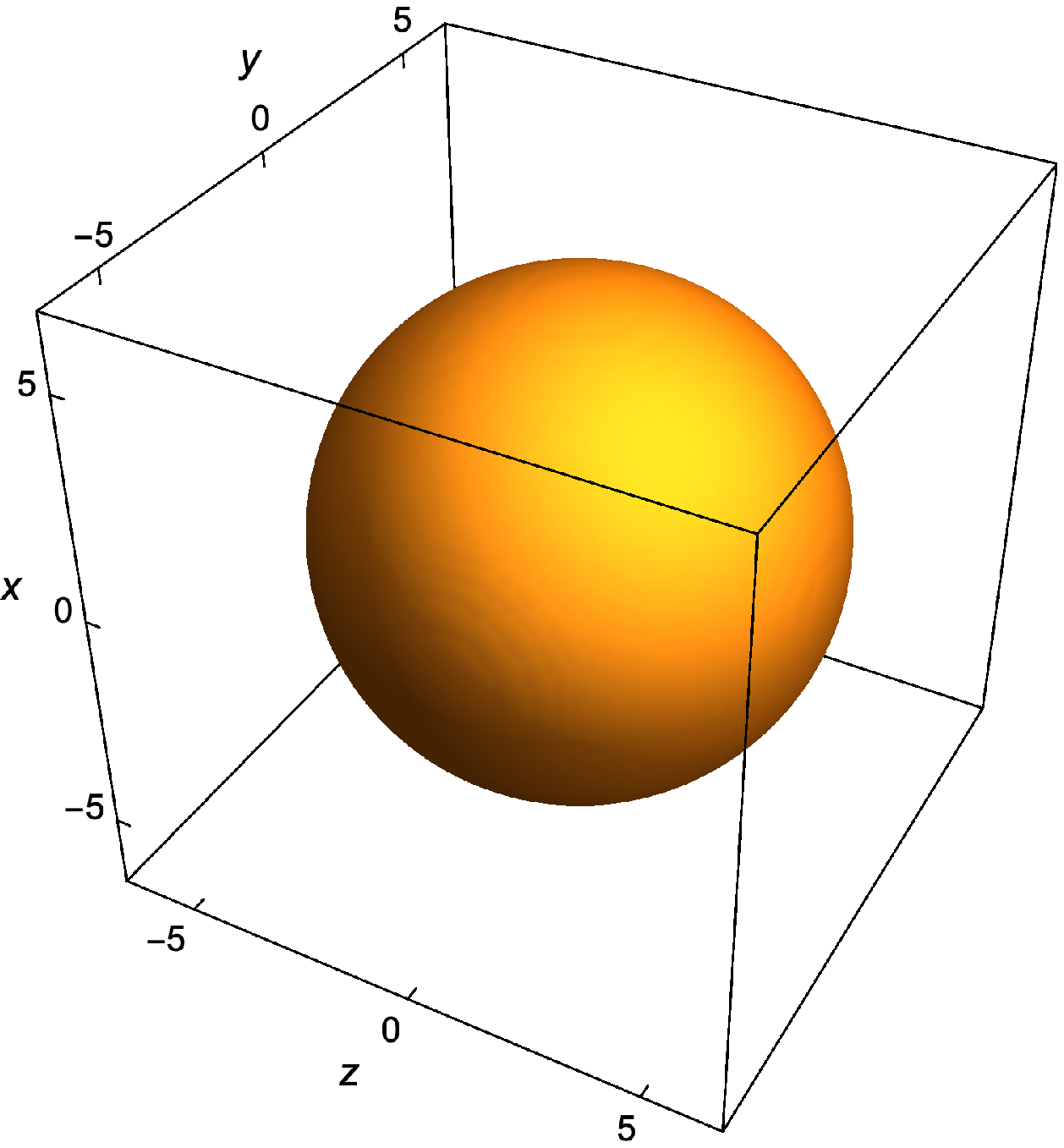} %
\includegraphics[width=0.35\textwidth]{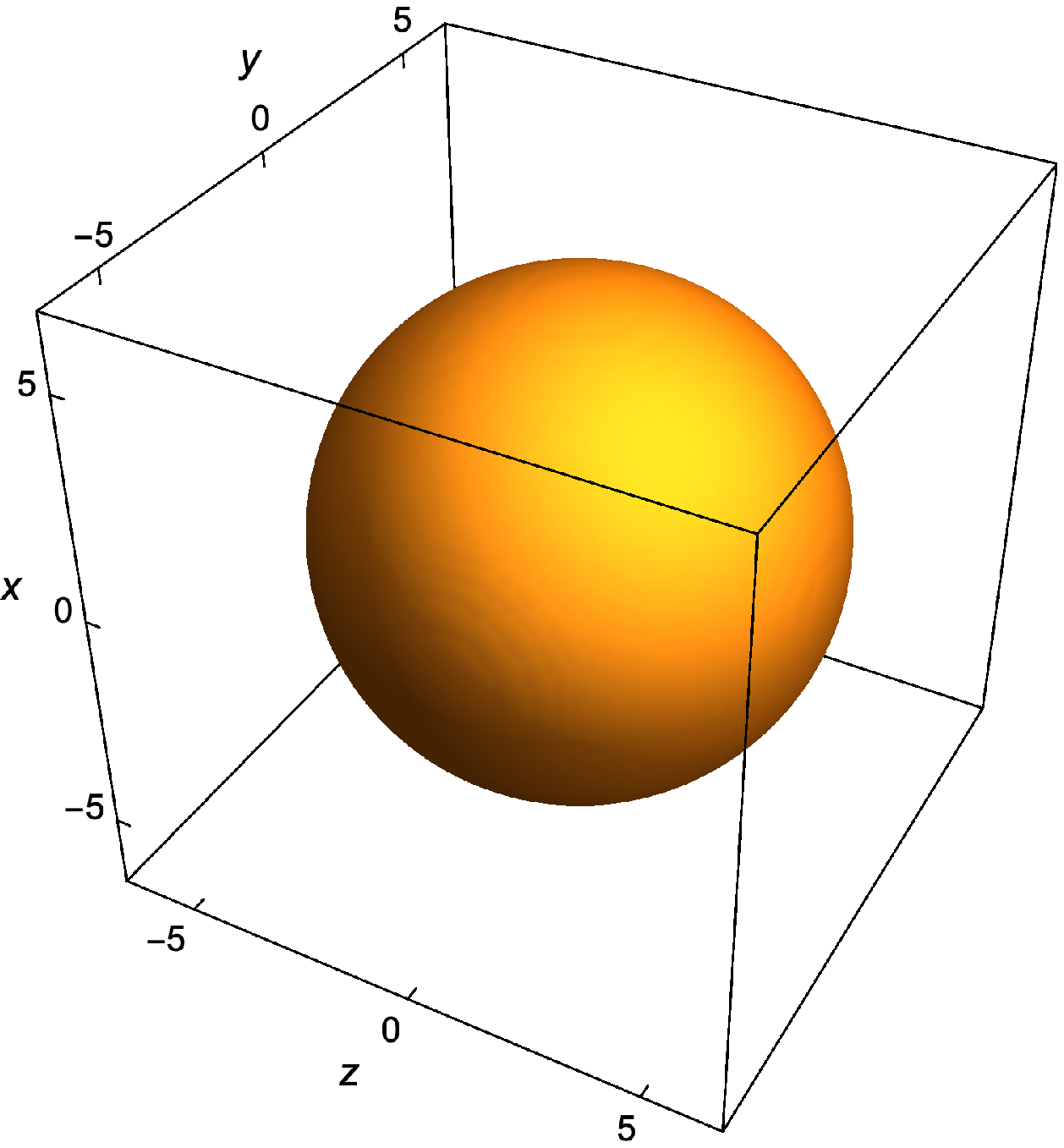}
\end{center}
\caption{ Stabilization of the three-dimensional $^{52}$Cr trapless dipolar BECs with $\protect\epsilon _{0}=$ $-25a_{0}$ and $\protect\epsilon _{1}=0$, $\protect\chi _{1}=0$. Panel (1,1) $\protect\chi _{0}=$ -0.013 $\protect\epsilon _{0}$ at $t=2\,$ms, panel (1,2) $\protect\chi _{0}=$ -0.012 $\protect\epsilon _{0}$ at $t=2.5\,$ms, panel (2,1) $\protect\chi _{0}=$ -0.011 $\protect\epsilon _{0}$ at $t=3\,$ms and panel (2,2) $\protect\chi _{0}=$ -0.01085 $\protect\epsilon _{0}$ at $t=5\,$ms. } \label{f6}
\end{figure}

In the course of time evolution, we increase the nonlinearity coefficients from $0$ at each time step as
$g(t)=f(t)g_f\{ a_1 - b_1\sin [\Omega \,(t-\tau)] \} $ and $\chi(t)=f(t)\chi_f\{ a_2 - b_2\sin [\Omega \,(t-\tau)] \} $, with $f(t)=t/\tau $ for $0\leq t\leq \tau $ and $f(t)=1$ for $t>\tau$. 
At the same time, the trap is removed by changing $d(t)$ from $1$ to $0$ by $d(t)=1-f(t)$. Thus, the trap is eventually removed and the final values of the coefficients are attained at $t=\tau $ and the periodically oscillating nonlinearities become $g(t)=g_{f}[a_{1}-b_{1}\sin \left(\Omega t\right) ]$ and $\chi(t)=\chi_{f}[a_{2}-b_{2}\sin \left(\Omega t\right) ]$ is introduced for $t>\tau $, cf. Refs. \cite{Sabari2010,Adhikari2004,Saito2003}.  After the complete removal of external trap, the condensates will collapse for strong attraction due to the final nonlinearity. If the attraction due to the final value of nonlinearity after switching off the external trap is too weak, the system becomes weakly attractive and the condensates start to expand. 

In Fig.~\ref{f6}, we present the results for the condensate of $^{52}$Cr atoms which have a moderate dipole moment corresponding to $a_{dd}=16a_{0}$~\cite{Koch:2008,Lahaye:2009}. The figure demonstrates how one can stabilize the condensate with $\protect\epsilon _{0}=$ $-25a_{0}$ by varying the strengths of the repulsive three-body  interactions (without the contribution of the oscillatory part of the interactions). In panel (1,1) the condensates are found to be stable up to $2ms$ for $\protect\chi _{0}=$ -0.013$\protect\epsilon_{0}$. Similarly, in panels  (1,2), (2,1) and (2,2), the condensates are stable up to $2.5ms$, $3ms$ and $5ms$ for $\protect\chi _{0}=$ -0.012$\protect\epsilon_{0}$, -0.011$\protect\epsilon_{0}$ and -0.01085$\protect\epsilon_{0}$, respectively. Further evolution of  the condensates may perhaps lead  to their collapse or expansion.

\section{Conclusion}

\label{sec5}

We have considered the dynamics of BEC with parameters corresponding to $^{52}$%
Cr with the motivation of studying the impact of the three-body interaction on the
stabilization of trapless dipolar BECs with two-body contact interactions.
By means of the variational approximation, we have produced phase diagrams
showing stable states of the trapless dipolar BECs which indicates the 
enhancement of the stabilization of the condensates by the reinforcement of  binary interaction with three-body
collisions. The stability is indicated by the presence of the minimum of the
effective potential and stable oscillation of the widths in the course of
the evolution. Further, we have performed full 3D simulations to confirm the
stability. In particular, the stability may be enhanced by the time-periodic modulation of the three-body interaction added to the attractive binary
interaction.

It may be relavant to extend the present analysis to  3D self-trapped  condensates with
embedded vorticity. In that case, stability of the solitary vortices against
spontaneous splitting is a crucially important issue \cite{splitting}.

\section*{Acknowledgements}

\noindent Authors wish to thank the referees for their critical comments to improve the focus of the paper. SS wishes to thank the Council of Scientific and Industrial
Research(CSIR) the Government of India, for financial assistance (Grant No.
03(1456)/19/EMR-II). RKK acknowledges support from Marsden Fund (Contract
UOO1726). RR wishes to acknowledge the financial assistance from
DAE-NBHM(Grant No.02011/3/20/2020/NBHM(R.P)/RD II) and CSIR(Grant No
03(1456)/19/EMR-II). The work of BAM was supported, in part, by the Israel Science Foundation, through grant No. 1286/17.


\begin{thebibliography}{999} 

\bibitem{Koch:2008} Koch, T.; Tobias, L.; Thierry, M.; Jonas, F.; Bernd, G.; Axel, P.A. Stabilization of a purely dipolar quantum gas against collapse. {\it Nat. Phys.} \textbf{2008}, 4, 218.

\bibitem{Lahaye:2009} Lahaye, T.;  Menotti, C.; Santos, L.; Lewenstein, M.; Pfau, T.; The physics of dipolar bosonic quantum gases. {\it Rep. Prog. Phys.} \textbf{2009}, 72, 126401.

\bibitem{Lu:2011} Lu, M.; Burdick, N.Q.; Youn, S.H.; Lev, B.L. Strongly Dipolar Bose-Einstein Condensate of Dysprosium. Lev, {\it Phys. Rev. Lett.} \textbf{2011}, 107, 190401.

\bibitem{Youn:2010} Youn, S.H.; Lu, M.; Ray, U.; Lev, B.L. Dysprosium magneto-optical traps. {\it Phys. Rev. A} \textbf{2010}, 82, 043425.

\bibitem{Aikawa:2012} Aikawa, K.; Frisch, A.; Mark, M.; Baier, S.; Rietzler, A.; Grimm, R.; Ferlaino, F. Bose-Einstein Condensation of Erbium. {\it Phys. Rev. Lett.} \textbf{2012}, 108, 210401.   


\bibitem{Baranov:2008} Baranov, M. Theoretical progress in many-body physics with ultracold dipolar gases. {\it Phys. Rep.} \textbf{2008}, 464, 71.

\bibitem{Santos:2003} Santos, L.; Shlyapnikov, G.V.; Lewenstein, M. Roton-Maxon Spectrum and Stability of Trapped Dipolar Bose-Einstein Condensates. {\it Phys. Rev. Lett.} \textbf{2003}, 90, 250403.

\bibitem{Goral:2002} Goral, K.; Santos, L.Ground state and elementary excitations of single and binary Bose-Einstein condensates of trapped dipolar gases. {\it Phys. Rev. A} \textbf{2002}, 66, 023613.

\bibitem{Wilson:2010} Wilson, R.M.; Ronen, S.; Bohn, J. L. Critical Superfluid Velocity in a Trapped Dipolar Gas. {\it Phys. Rev. Lett.} \textbf{2010}, 104, 094501. 

\bibitem{Ticknor:2011} Ticknor, C.; Wilson, R.M.; Bohn, J.L. Anisotropic Superfluidity in a Dipolar Bose Gas. {\it Phys. Rev. Lett.} \textbf{2011}, 106, 065301.


\bibitem{Tieleman:2011} O. Tieleman, A. Lazarides, and C. Morais Smith, Supersolid phases of dipolar bosons in optical lattices with a staggered flux. {\it Phys. Rev. A} \textbf{2011}, 83, 013627.

\bibitem{Zhou:2010} K. Zhou, Z. Liang, and Z. Zhang, Quantum phases of a dipolar Bose-Einstein condensate in an optical lattice with three-body interaction. {\it Phys. Rev. A} \textbf{2010}, 82, 013634.

\bibitem{vor1} Mulkerin, B.C.; van Bijnen, R.M.W.; O'Dell, D.H.J.; Martin, A.M.; Parker, N.G.  Anisotropic and Long-Range Vortex Interactions in Two-Dimensional Dipolar Bose Gases. {\it Phys. Rev. Lett.} \textbf{2013}, 111, 170402.

\bibitem{vor2} Martin, A.M.; Marchant, N.G.; O'Dell, D.H.J.; Parker, N.G.  Vortices and vortex lattices in quantum ferrofluids. {\it J. Phys. Cond. Matt.} \textbf{2017}, 29, 103004.

\bibitem{Sabari2017hv} Sabari, S. Vortex formation and hidden vortices in dipolar Bose–Einstein condensates. {\it Phys. Lett. A} \textbf{2017}, 381, 3062.

\bibitem{Sabari2018a} Sabari, S.; Kishor Kumar, R. Effect of an oscillating Gaussian obstacle in a dipolar Bose-Einstein condensate. {\it Eur. Phys. J. D} \textbf{2018}, 72, 48.

\bibitem{Tikhonenkov:2008} Tikhonenkov, I.; Malomed, B.A.; Vardi, A. Anisotropic Solitons in Dipolar Bose-Einstein Condensates. {\it Phys. Rev. Lett.} \textbf{2008}, 100, 090406.

\bibitem{Koberle} K\"{o}berle, P.; Zajec, D.; Wunner, G.; Malomed, B.A. Creating two-dimensional bright solitons in dipolar Bose-Einstein condensates. {\it Phys. Rev. A} \textbf{2012}, 85, 023630.

\bibitem{Pfau1} Ferrier-Barbut, I.; Kadau, H.; Schmitt, M.; Wenzel, M.; Pfau, T. Observation of Quantum Droplets in a Strongly Dipolar Bose Gas. {\it Phys. Rev. Lett.} \textbf{2016}, 116, 215301.

\bibitem{Pfau3} Schmitt, M.; Wenzel, M.; B\"{u}ttcher, F.; Ferrier-Barbut, I.; Pfau, T. Self-bound droplets of a dilute magnetic quantum liquid. {\it Nature} \textbf{2016}, 539, 259.

\bibitem{Santos1} Chomaz, L.; Baier, S.; Petter, D.; Mark, M.J.; Wachtler, F.; Santos, L.; Ferlaino, F.  Quantum-Fluctuation-Driven Crossover from a Dilute Bose-Einstein Condensate to a Macrodroplet in a Dipolar Quantum Fluid. {\it Phys. Rev. X} \textbf{2016}, 6, 041039.

\bibitem{competing} Cuevas, J.; Malomed, B.A.; Kevrekidis, P.G.; Frantzeskakis, D.J. Solitons in quasi-one-dimensional Bose-Einstein condensates with competing dipolar and local interactions. {\it Phys. Rev. A} \textbf{2009}, 79, 053608.










\bibitem{Muruganandam:2012} Muruganandam, P.; Adhikari, S.K. Numerical and variational solutions of the dipolar Gross-Pitaevskii equation in reduced dimensions. {\it Las. Phys.} \textbf{2012}, 22, 813.

\bibitem{Lahaye:2008} Lahaye, T.; Metz, J.; Frohlich, B.;  Koch, T.; Meister, M.; Griesmaier, A. ; Pfau, T.; Saito, H.; Kawaguchi, Y.; Ueda,  M. \textbf{\it d}-Wave Collapse and Explosion of a Dipolar Bose-Einstein Condensate. {\it Phys. Rev. Lett.} \textbf{2008}, 101, 080401.

\bibitem{RKK:2015} Kishor Kumar, R.; Young-S., L.E.; Vudragovi\'{c}, D.; Bala\v{z}, A.; Muruganandam, P.; Adhikari, S.K. Fortran and C programs for the time-dependent dipolar Gross–Pitaevskii equation in an anisotropic trap. {\it Comput. Phys. Commun.} \textbf{2015}, 195, 117.

\bibitem{Strecker:2002} Strecker, K.E.; Partridge, G. B.; Truscott, A. G.; Hulet, R. G. Formation and propagation of matter-wave soliton trains. {\it Nature} \textbf{417}, 150 (2002).

\bibitem{Khaykovich:1995} Khaykovich, L.; Schreck, F.; Ferrari, G.; Bourdel, T.; Cubizolles, J.; Carr, L.; Castin, Y.; Salomon, C. Formation of a Matter-Wave Bright Soliton. {\it Science} \textbf{296}, 1290 (2002).

\bibitem{Krolikowski:2001} Krolikowski, W.; Bang, O.; Rasmussen, J.J.;  Wyller, J. Modulational instability in nonlocal nonlinear Kerr media. {\it Phys. Rev. E} \textbf{64}, 016612 (2001).

\bibitem{Bang:2002} Bang, O.; Krolikowski, W.; Wyller, J.;  Rasmussen, J.J. Collapse arrest and soliton stabilization in nonlocal nonlinear media. {\it Phys. Rev. E} \textbf{66}, 046619 (2002).

\bibitem{Landau:1960} Landau L.; Lifshitz,   E. M. \textit{Mechanics}. {\it Pergamon, Oxford, Oxford} \textbf{1960}.

\bibitem{Towers2002} Towers, I.; Malomed, B.A. Stable (2+1)-dimensional solitons in a layered medium with sign-alternating Kerr nonlinearity. {\it J. Opt. Soc. Am. B} \textbf{2002}, 19, 537.

\bibitem{Abdullaev2003} 
Abdullaev, F.K,; Caputo, J.G; Kraenkel, R.A.; Malomed, B.A.. Controlling collapse in Bose-Einstein condensates by temporal modulation of the scattering length. {\it Phys. Rev. A} \textbf{2003}, 67, 013605.

\bibitem{Saito2003} 
Saito, H.; Ueda, M. Dynamically Stabilized Bright Solitons in a Two-Dimensional Bose-Einstein Condensate. {\it Phys. Rev. Lett.} \textbf{2003}, 90, 040403.

\bibitem{Adhikari2004} 
Adhikari, S.K. Stabilization of bright solitons and vortex solitons in a trapless three-dimensional Bose-Einstein condensate by temporal modulation of the scattering length.  {\it Phys. Rev. A} \textbf{2004}, 69, 063613.


\bibitem{Sabari2010} 
Sabari, S.; Raja, R.V.J.; Porsezian, K.; Muruganandam, P. Stability of trapless Bose–Einstein condensates with two-and three-body interactions. {\it J. Phys. B: At. Mol. Opt. Phys.} \textbf{2010}, 43, 125302.

\bibitem{Sabari2017} 
Sabari, S.; Porsezian, K.; Muruganandam, P. Dynamical stabilization of two-dimensional trapless Bose–Einstein condensates by three-body interaction and quantum fluctuations. {\it Chaos, Solitons and Fractals.} \textbf{2017}, 103, 232.

\bibitem{Tamil} Tamil Thiruvalluvar, R.; Sabari, S.; Porsezian, K. Stabilization of repulsive trapless Bose–Einstein condensates. {\it J. Phys. B: At. Mol. Opt. Phys.} \textbf{2018}, 51, 165202.

\bibitem{book} Malomed, B.A.;, \textit{Soliton Management in Periodic Systems}. {\it Springer: New York.} \textbf{2006}.

\bibitem{Kivshar} Kivshar, Y.S.; Agrawal, G.P. \textit{Optical Solitons: From Fibers to Photonic Crystals}. {\it Academic Press San Diego} \textbf{2003}.

\bibitem{Kartashov} Kartashov, Y.V.; Malomed, B.A.; Torner, L. Solitons in nonlinear lattices. {\it Rev. Mod. Phys.} \textbf{2011}, 83, 247.

\bibitem{Zeng} Zeng, J.; Malomed, B.A. Stabilization of one-dimensional solitons against the critical collapse by quintic nonlinear lattices. {\it Phys. Rev. A} \textbf{2012}, 85, 023824.


\bibitem{Sakaguchi} Sakaguchi, H.; Malomed, B.A. Matter-wave solitons in nonlinear optical lattices. {\it Phys. Rev. E} \textbf{2005}, 72 046610.

\bibitem{Abdullaev} Abdullaev F.Kh.;  Garnier, J. Propagation of matter-wave solitons in periodic and random nonlinear potentials. {\it Phys. Rev. A} \textbf{2005}, 72 061605(R).

\bibitem{Abdullaev2003a} Abdullaev, F.Kh.; Kamchatnov, A.M.; Konotop, V.V; Brazhnyi, V.A. Adiabatic Dynamics of Periodic Waves in Bose-Einstein Condensates with Time Dependent Atomic Scattering Length. {\it Phys. Rev. Lett.} \textbf{2003}, 90, 230402.

\bibitem{Garcia} Perez-Garcia, V.; Konotop, V.; Brazhnyi, V.A. Feshbach Resonance Induced Shock Waves in Bose-Einstein Condensates. {\it Phys. Rev Lett.} \textbf{2004}, 92, 220403.

\bibitem{Pacciani} Konotop, V.V.; Pacciani, P. Collapse of Solutions of the Nonlinear Schrödinger Equation with a Time-Dependent Nonlinearity: Application to Bose-Einstein Condensates. {\it Phys. Rev Lett.} \textbf{2005}, 94, 240405.


\bibitem{Belmonte1} Belmonte-Beitia, J.;  P\'{e}rez-Garc\'{\i}a V.M.; Vekslerchik, V.; Torres, P. J. Localized Nonlinear Waves in Systems with Time- and Space-Modulated Nonlinearities. {\it Phys. Rev. Lett.} \textbf{2008}, 100, 164102.

\bibitem{Deng} Wang, D.-S.; Hu, X.-H.; Liu, W.M. Localized nonlinear matter waves in two-component Bose-Einstein condensates with time- and space-modulated nonlinearities. {\it Phys. Rev. A} \textbf{2010}, 82, 023612.

\bibitem{Sabari2015} Sabari, S.; Jisha, C.P.; Porsezian, K.; Brazhnyi, V.A. Dynamical stability of dipolar Bose-Einstein condensates with temporal modulation of the s-wave scattering length. {\it Phys. Rev. E} \textbf{2015}, 92, 032905.

\bibitem{sabari2019}Tamilthiruvalluvar, R.;  Sabari, S. Stabilization of trapless Bose– Einstein condensates without any management. Phys. Lett. A \textbf{2019}, 383, 2033.

\bibitem{Sabari2018b} Sabari, S.; Biswajyoti Dey. Stabilization of trapless dipolar Bose-Einstein condensates by temporal modulation of the contact interaction. {\it Phys. Rev. E} \textbf{2018}, 98, 042203.


\bibitem{Keltoum:2019} Keltoum, R.; Boudjemaa, A. Dipolar Bose gas with three-body interactions in weak disorder. {\it Eur. Phys. J. D} \textbf{2019}, 73, 115.

\bibitem{Boudjemaa:2018} Boudjemaa, A. 
Dipolar Bose gas with three-body interactions at finite temperature. {\it J. Phys. B: At. Mol. Opt. Phys.} \textbf{2018}, 51, 025203.

\bibitem{Blakie:2016} Blakie, P.B. Properties of a dipolar condensate with three-body interactions. {\it Phys. Rev. A} \textbf{2016}, 93, 033644.

\bibitem{Xi:2016} Xi, K.-T.; Saito, H. Droplet formation in a Bose-Einstein condensate with strong dipole-dipole interaction. {\it Phys. Rev. A} \textbf{2016}, 93, 011604(R).

\bibitem{Bisset:2015} Bisset, R.N.; Blakie, P. B. Crystallization of a dilute atomic dipolar condensate. {\it Phys. Rev. A} \textbf{2015}, 92, 061603(R).

\bibitem{Lu:2015} Lu, Z.-K.; Yun Li, Petrov, D.S.; Shlyapnikov, G.V. Stable Dilute Supersolid of Two-Dimensional Dipolar Bosons. {\it Phys. Rev. Lett.} \textbf{2015}, 115, 075303.

\bibitem{3body1} Bulgac, A. Dilute Quantum Droplets. {\it Phys. Rev. Lett.} \textbf{2002}, 89, 050402.

\bibitem{3body2}Braaten, E., Hammer, H.W. Mehen, T. Dilute Bose-Einstein Condensate with Large Scattering Length. {\it Phys. Rev. Lett.} \textbf{2002}, 88, 040401.

\bibitem{Gammal2000}Gammal, A.; Frederico, T.; Tomio, L.; Chomaz, Ph. Atomic Bose-Einstein condensation with three-body interactions and collective excitations. {\it J. Phys. B: At. Mol. Opt. Phys.} \textbf{2000}, 33, 4053.

\bibitem{Fibich} Fibich, G. \textit{The nonlinear Schr\"{o}dinger equation: singular solutions and optical collapse}. {\it Heidelberg: Springer}. \textbf{2015}.

\bibitem{splitting} Mihalache, D.; Mazilu, D.;  Crasovan, L.-C.; Towers, I.; Malomed, B.A.; Buryak, A.V.; Torner, L.; Lederer,  F. Stable three-dimensional spinning optical solitons supported by competing quadratic and cubic nonlinearities. {\it Phys. Rev. E} \textbf{2002}, 66, 016613.

\end{thebibliography}
\end{document}